\documentclass[aps,prd,preprint,superscriptaddress,tightenlines,%
nofootinbib]{revtex4}
\newcommand{\PRE}[1]{{#1}}   

\usepackage{bm}
\usepackage{epsfig}     
\usepackage{latexsym}

\newcommand{\postscript}[2]{\setlength{\epsfxsize}{#2\hsize}
   \centerline{\epsfbox{#1}}}
\newcommand{\comment}[1]{}

\def\comment#1{{}}

\def\tfrac#1#2{{\textstyle\frac{#1}{#2}}}
\def\slashchar#1{\setbox0=\hbox{$#1$}           
   \dimen0=\wd0                                 
   \setbox1=\hbox{/} \dimen1=\wd1               
   \ifdim\dimen0>\dimen1                        
      \rlap{\hbox to \dimen0{\hfil/\hfil}}      
      #1                                        
   \else                                        
      \rlap{\hbox to \dimen1{\hfil$#1$\hfil}}   
      /                                         
   \fi}
\newif\ifnref

\nreffalse

\input epsf
\def\figin{\epsfcheck\figin}\def\figins{\epsfcheck\figins}
\def\epsfcheck{\ifx\epsfbox\UnDeFiNeD
\message{(NO epsf.tex, FIGURES WILL BE IGNORED)}
\gdef\figin##1{\vskip2in}\gdef\figins##1{\hskip.5in}
\else\message{(FIGURES WILL BE INCLUDED)}%
\gdef\figin##1{##1}\gdef\figins##1{##1}\fi}
\def\DefWarn#1{}
\def\figinsert{\goodbreak\midinsert}
\def\ifig#1#2#3{\DefWarn#1\xdef#1{fig.~\the\figno}
\writedef{#1\leftbracket fig.\noexpand~\the\figno}%
\figinsert\figin{\centerline{#3}}\medskip\centerline{\vbox{\baselineskip12pt
\advance\hsize by -1truein\noindent\footnotefont{\bf Fig.~\the\figno } #2}}
\bigskip\endinsert\global\advance\figno by1}

\def\hat{\widehat}

\begin{document}

\preprint{
\hfil
\begin{minipage}[t]{3in}
\begin{flushright}
\vspace*{.4in}
MPP--2007--182\\
NUB-Th-3261
\end{flushright}
\end{minipage}
}

\title{Jet signals for low mass strings at the LHC
\PRE{\vspace*{0.3in}} }

\author{Luis A. Anchordoqui}
\affiliation{Department of Physics,\\
University of Wisconsin-Milwaukee,
 Milwaukee, WI 53201, USA
\PRE{\vspace*{.1in}}
}

\author{Haim Goldberg}
\affiliation{Department of Physics,\\
Northeastern University, Boston, MA 02115, USA
\PRE{\vspace*{.1in}}
}

\author{Satoshi Nawata}
\affiliation{Department of Physics,\\
University of Wisconsin-Milwaukee,
 Milwaukee, WI 53201, USA
\PRE{\vspace*{.1in}}
}

\author{Tomasz R. Taylor}
\affiliation{Department of Physics,\\
Northeastern University, Boston, MA 02115, USA
\PRE{\vspace*{.1in}}
}

\affiliation{Max--Planck--Institut f\"ur Physik\\
 Werner--Heisenberg--Institut,
80805 M\"unchen, Germany
\PRE{\vspace*{.1in}}
}

\date{December 2007}
\PRE{\vspace*{.5in}}
\begin{abstract}\vskip 3mm\noindent
  The mass scale $M_s$ of superstring theory is an arbitrary parameter
  that can be as low as few TeVs if the Universe contains large extra
  dimensions. We propose a search for the effects of Regge excitations
  of fundamental strings at LHC, in the process $pp \to \gamma$ +
  jet. The underlying parton process is dominantly the single photon
  production in gluon fusion, $gg \to \gamma g$, with open
  string states propagating in intermediate channels. If the photon
  mixes with the gauge boson of the baryon number, which is a common
  feature of D-brane quivers, the amplitude appears already at the
  string disk level. It is completely determined by the mixing
  parameter -- and it is otherwise model-(compactification-)
  independent. Even for relatively small mixing, 100 fb$^{-1}$ of LHC
  data could probe deviations from standard model physics, at a
  $5\sigma$ significance, for $M_s$ as large as 3.3~TeV.
\end{abstract}

\maketitle

At the time of its formulation and for years thereafter, Superstring
Theory was regarded as a unifying framework for Planck-scale quantum
gravity and TeV-scale Standard Model (SM) physics.\comment{The theory
  lived in ten dimensions, so that a compactification scheme was
  necessary to provide a link to our 4D world.} Important advances
were fueled by the realization of the vital role played by
D-branes~\cite{joe} in connecting string theory to
phenomenology~\cite{reviews}.  This has permitted the formulation of
string theories with compositeness setting in at TeV
scales~\cite{Antoniadis:1998ig} and large extra dimensions. There are
two paramount phenomenological consequences for TeV scale D-brane
string physics: the emergence of Regge recurrences at parton collision
energies $\sqrt{\hat s} \sim {\rm string\ scale} \equiv M_s;$ and the
presence of one or more additional $U(1)$ gauge symmetries, beyond the
$U(1)_Y$ of the SM. The latter follows from the property that the
gauge group for open strings terminating on a stack of $N$ identical
D-branes is $U(N)$ rather than $SU(N)$ for $N>2.$ (For $N=2$ the gauge
group can be $Sp(1)$ rather than $U(2)$.) In this Letter we exploit
both these properties in order to obtain a ``new physics'' signal at
LHC which, if traced to low scale string theory, could with 100
fb$^{-1}$ of integrated luminosity probe deviations from SM physics at
a $5\sigma$ significance for $M_s$ as large as 3.3~TeV.

To develop our program in the simplest way, we will work within the
construct of a minimal model in which we consider scattering processes
which take place on the (color) $U(3)$ stack of D-branes. In the
bosonic sector, the open strings terminating on this stack contain, in
addition to the $SU(3)$ octet of gluons, an extra $U(1)$ boson
($C_\mu$, in the notation of~\cite{Berenstein:2006pk}), most simply
the manifestation of a gauged baryon number symmetry. The $U(1)_Y$
boson $Y_\mu$, which gauges the usual electroweak hypercharge
symmetry, is a linear combination of $C_\mu$, the $U(1)$ boson $B_\mu$
terminating on a separate $U(1)$ brane, and perhaps a third additional
$U(1)$ (say $W_\mu$) sharing a $U(2)$ brane which is also a
terminus for the $SU(2)_L$ electroweak gauge bosons $W_\mu^a.$ Thus,
critically for our purposes, the photon $A_\mu$, which is a linear
combination of $Y_\mu$ and $W^3_\mu$ {\em will participate with the
  gluon octet in (string) tree level scattering processes on the color
  brane, processes which in the SM occur only at one-loop level.} Such
a mixing between hypercharge and baryon number is a generic property
of D-brane quivers, see {\it e.g}.\
Refs.\cite{ant,bo,Berenstein:2006pk}.

The process we consider (at the parton level) is $gg\rightarrow
g\gamma$, where $g$ is an $SU(3)$ gluon and $\gamma$ is the photon. As
explicitly calculated below, this will occur at string disk (tree)
level, and will be manifest at LHC as a non-SM contribution to
$pp\rightarrow \gamma +\ {\rm jet}$.  A very important property of
string disk amplitudes is that they are completely model-independent;
thus the results presented below are robust, because {\em they hold
  for arbitrary compactifications of superstring theory from ten to four
  dimensions, including those that break supersymmetry}.  The SM
background for this signal originates in the parton tree level
processes $g q \rightarrow \gamma q,\ g\bar q\rightarrow \gamma\bar q\
{\rm and} \ q\bar q\rightarrow \gamma g$. Of course, the SM processes
will also receive stringy corrections which should be added to the
pure bosonic contribution as part of the
signal~\cite{Cullen:2000ef,Burikham:2004su,Meade:2007sz,Domokos:1998ry}. 
We leave this evaluation to
a subsequent publication~\cite{ws}; thus, the contribution from the
bosonic process calculated here is to be regarded as a lower bound to
the stringy signal. It should also be stated that, in what follows, we
do not include effects of Kaluza-Klein recurrences due to
compactification. We assume that all such effects are in the
gravitational sector, and hence occur at higher order in string
coupling~\cite{Cullen:2000ef}.

The most direct way to compute the amplitude for the scattering of
four gauge bosons is to consider the case of polarized particles
because all non-vanishing contributions can be then generated from a
single, maximally helicity violating (MHV), amplitude -- the so-called
{\it partial\/} MHV amplitude~\cite{ptmhv}.  Assume that two vector bosons,
with the momenta $k_1$ and $k_2$, in the $U(N)$ gauge group states
corresponding to the generators $T^{a_1}$ and $T^{a_2}$ (here in the
fundamental representation), carry negative helicities while the other
two, with the momenta $k_3$ and $k_4$ and gauge group states $T^{a_3}$
and $T^{a_4}$, respectively, carry positive helicities.  Then the
partial amplitude for such an MHV configuration is given by~\cite{STi,STii}
\begin{equation}
\label{ampl}
A(1^-,2^-,3^+,4^+) ~=~ 4\, g^2\, {\rm Tr}
  \, (\, T^{a_1}T^{a_2}T^{a_3}T^{a_4}) {\langle 12\rangle^4\over
    \langle 12\rangle\langle 23\rangle\langle 34\rangle\langle
    41\rangle}V(k_1,k_2,k_3,k_4)\ ,
\end{equation}
where $g$ is the $U(N)$ coupling constant, $\langle ij\rangle$ are the standard
spinor products written in the notation of
Refs.~\cite{Mangano,Dixon}, and the Veneziano formfactor,
\begin{equation}
\label{formf}
V(k_1,k_2,k_3,k_4)=V(s,t,u)= {\Gamma(1-s)\ \Gamma(1-u)\over
    \Gamma(1+t)}\ ,
\end{equation}
is the function of Mandelstam variables, here
normalized in the string units:
\begin{equation}
\label{mandel}
s={2k_1k_2\over M_s^2},~ t={2
  k_1k_3\over M_s^2}, ~u={2 k_1k_4 \over M_s^2}:\qquad s+t+u=0.
\end{equation}
(For simplicity we drop carets for the parton subprocess.)
Its low-energy
  expansion reads
\begin{equation}
\label{vexp}
V(s,t,u)\approx 1-{\pi^2\over 6}s\,
    u-\zeta(3)\,s\, t\, u+\dots
\end{equation}

We are interested in the amplitude involving three $SU(N)$ gluons
$g_1,~g_2,~g_3$ and one $U(1)$ gauge boson $\gamma_4$ associated to
the same $U(N)$ quiver:
\begin{equation}
\label{gens}
T^{a_1}=T^a \ ,~ \ T^{a_2}=T^b\ ,~ \
  T^{a_3}=T^c \ ,~ \ T^{a_4}=QI\ ,
\end{equation}
where $I$ is the $N{\times}N$ identity matrix and $Q$ is the
$U(1)$ charge of the fundamental representation. The $U(N)$
generators are normalized according to
\begin{equation}
\label{norm}
{\rm Tr}(T^{a}T^{b})={1\over 2}\delta^{ab}.
\end{equation}
Then the color
factor \begin{equation}\label{colf}{\rm
    Tr}(T^{a_1}T^{a_2}T^{a_3}T^{a_4})=Q(d^{abc}+{i\over 4}f^{abc})\ ,
\end{equation}
where the totally symmetric symbol $d^{abc}$ is the symmetrized trace
while $f^{abc}$ is the totally antisymmetric structure constant.

The full MHV amplitude can be obtained~\cite{STi,STii} by summing
the partial amplitudes (\ref{ampl}) with the indices permuted in the
following way: \begin{equation}
\label{afull} {\cal M}(g^-_1,g^-_2,g^+_3,\gamma^+_4)
  =4\,g^{2}\langle 12\rangle^4 \sum_{\sigma } { {\rm Tr} \, (\,
    T^{a_{1_{\sigma}}}T^{a_{2_{\sigma}}}T^{a_{3_{\sigma}}}T^{a_{4}})\
    V(k_{1_{\sigma}},k_{2_{\sigma}},k_{3_{\sigma}},k_{4})\over\langle
    1_{\sigma}2_{\sigma} \rangle\langle
    2_{\sigma}3_{\sigma}\rangle\langle 3_{\sigma}4\rangle \langle
    41_{\sigma}\rangle }\ ,
\end{equation}
  where the sum runs over all 6
permutations $\sigma$ of $\{1,2,3\}$ and
$i_{\sigma}\equiv\sigma(i)$. As a result, the antisymmetric part of
the color factor (\ref{colf}) cancels and one obtains
\begin{equation}\label{mhva}
{\cal
    M}(g^-_1,g^-_2,g^+_3,\gamma^+_4)=8\, Q\, d^{abc}g^{2}\langle
  12\rangle^4\left({\mu(s,t,u)\over\langle 12\rangle\langle
      23\rangle\langle 34\rangle\langle
      41\rangle}+{\mu(s,u,t)\over\langle 12\rangle\langle
      24\rangle\langle 13\rangle\langle 34\rangle}\right),
\end{equation}
 where
\begin{equation}
\label{mudef}
\mu(s,t,u)= \Gamma(1-u)\left( {\Gamma(1-s)\over
      \Gamma(1+t)}-{\Gamma(1-t)\over \Gamma(1+s)}\right) .
\end{equation}
All
non-vanishing amplitudes can be obtained in a similar way. In
particular,
\begin{equation}
\label{mhvb}
{\cal M}(g^-_1,g^+_2,g^-_3,\gamma^+_4)=8\, Q\,
  d^{abc}g^{2}\langle 13\rangle^4\left({\mu(t,s,u)\over\langle
      13\rangle\langle 24\rangle\langle 14\rangle\langle
      23\rangle}+{\mu(t,u,s)\over\langle 13\rangle\langle
      24\rangle\langle 12\rangle\langle 34\rangle}\right),
\end{equation}
and the
remaining ones can be obtained either by appropriate permutations or
by complex conjugation.

In order to obtain the cross section for the (unpolarized) partonic
subprocess $gg\to g\gamma$, we take the squared moduli of individual
amplitudes, sum over final polarizations and colors, and average over
initial polarizations and colors. As an example, the modulus square of
the amplitude (\ref{afull}) is:
\begin{equation}
\label{mhvsq}
|{\cal
    M}(g^-_1,g^-_2,g^+_3,\gamma^+_4)|^2=64\, Q^2\, d^{abc}d^{abc}g^{4}
  \left|{s\mu(s,t,u)\over u}+{s\mu(s,u,t)\over t} \right|^2 \, .
\end{equation}
 Taking
into account all $4(N^2-1)^2$ possible initial polarization/color
configurations and the formula~\cite{groupf}
\begin{equation}
\label{dsq}
\sum_{a,b,c}d^{abc}d^{abc}={(N^2-1)(N^2-4)\over 16 N},
\end{equation}
 we
obtain the average squared amplitude
\begin{equation}
\label{mhvav}
|{\cal M}(gg\to
  g\gamma)|^2= g^4Q^2C(N)\left\{ \left|{s\mu(s,t,u)\over
        u}+{s\mu(s,u,t)\over t} \right|^2+(s\leftrightarrow
    t)+(s\leftrightarrow u)\right\},
\end{equation}
 where
\begin{equation}\label{cnn}
C(N)={2 (N^2-4)\over N(N^2-1)}.
\end{equation}

The two most interesting energy regimes of $gg\to g\gamma$
scattering are far below the string mass scale $M_s$
and near the threshold for the production of massive string
excitations. At low energies, Eq.~(\ref{mhvav}) becomes
\begin{equation}
\label{mhvlow}
|{\cal M}(gg\to
  g\gamma)|^2\approx  g^4Q^2C(N){\pi^4\over 4}(s^4+t^4+u^4)\qquad
  (s,t,u\ll 1) \, .
\end{equation}
The absence of massless poles, at $s=0$ {\it etc.\/}, translated
into the terms of effective field theory, confirms that there are
no exchanges of massless particles contributing to this process.
On the other hand, near the string threshold $s\approx M_s^2$
(where we now restore the string scale)
\begin{equation}
\label{mhvlow2}
|{\cal M}(gg\to g\gamma)|^2\approx
4 g^4Q^2C(N){M_s^8+t^4+u^4\over M_s^4[(s-M_s^2)^2+(\Gamma M_s)^2]}
\qquad (s\approx M_s^2),
\end{equation}
with the singularity (smeared with a width $\Gamma$) reflecting the
presence of a massive string mode propagating in the $s$ channel. In
what follows we will take $N=3$, set $g$ equal to the QCD coupling
constant $(g^2/4\pi = 0.1)$, and $\Gamma\simeq (g^2/16\pi) \, (2j
+1)^{-1} \, M_s$, with $j = 2$~\cite{nota}. Before proceeding with
numerical calculation, we need to make precise the value of $Q$. If we
were considering the process $gg\rightarrow C g,$ where $C$ is the
U(1) gauge field tied to the $U(3)$ brane, then $Q = \sqrt{1/6}$ due
to the normalization condition~(\ref{norm}). However, for
$gg\rightarrow \gamma g$ there are two additional projections: from
$C_\mu$ to the hypercharge boson $Y_\mu$, giving a mixing factor
$\kappa$; and from $Y_\mu$ onto a photon, providing an additional
factor $\cos\theta_W \ (\theta_W=$ Weinberg angle). The $C-Y$ mixing
coefficient is model dependent: in the minimal
model~\cite{Berenstein:2006pk} it is quite small, around $\kappa
\simeq 0.12$ for couplings evaluated at the $Z$ mass, which is
modestly enhanced to $\kappa \simeq 0.14$ as a result of RG running of
the couplings up to 2.5~TeV.  It should be noted that in
models~\cite{ant,bo} possessing an additional $U(1)$ which partners
$SU(2)_L$ on a $U(2)$ brane, the various assignment of the charges can
result in values of $\kappa$ which can differ considerably from
$0.12.$ In what follows, we take as a fiducial value $\kappa^2 =
0.02.$ Thus, if (\ref{mhvlow2}) is to describe $gg\rightarrow \gamma
g,$ \
\begin{equation}
Q^2= \tfrac{1}{6} \ \kappa^2 \ \cos^2\theta_W \simeq 2.55\times
10^{-3}\ \left(\kappa^2/0.02\right)\ \ .
\label{Q2}
\end{equation}

\begin{figure}[tbp]
\begin{minipage}[t]{0.49\textwidth}
\postscript{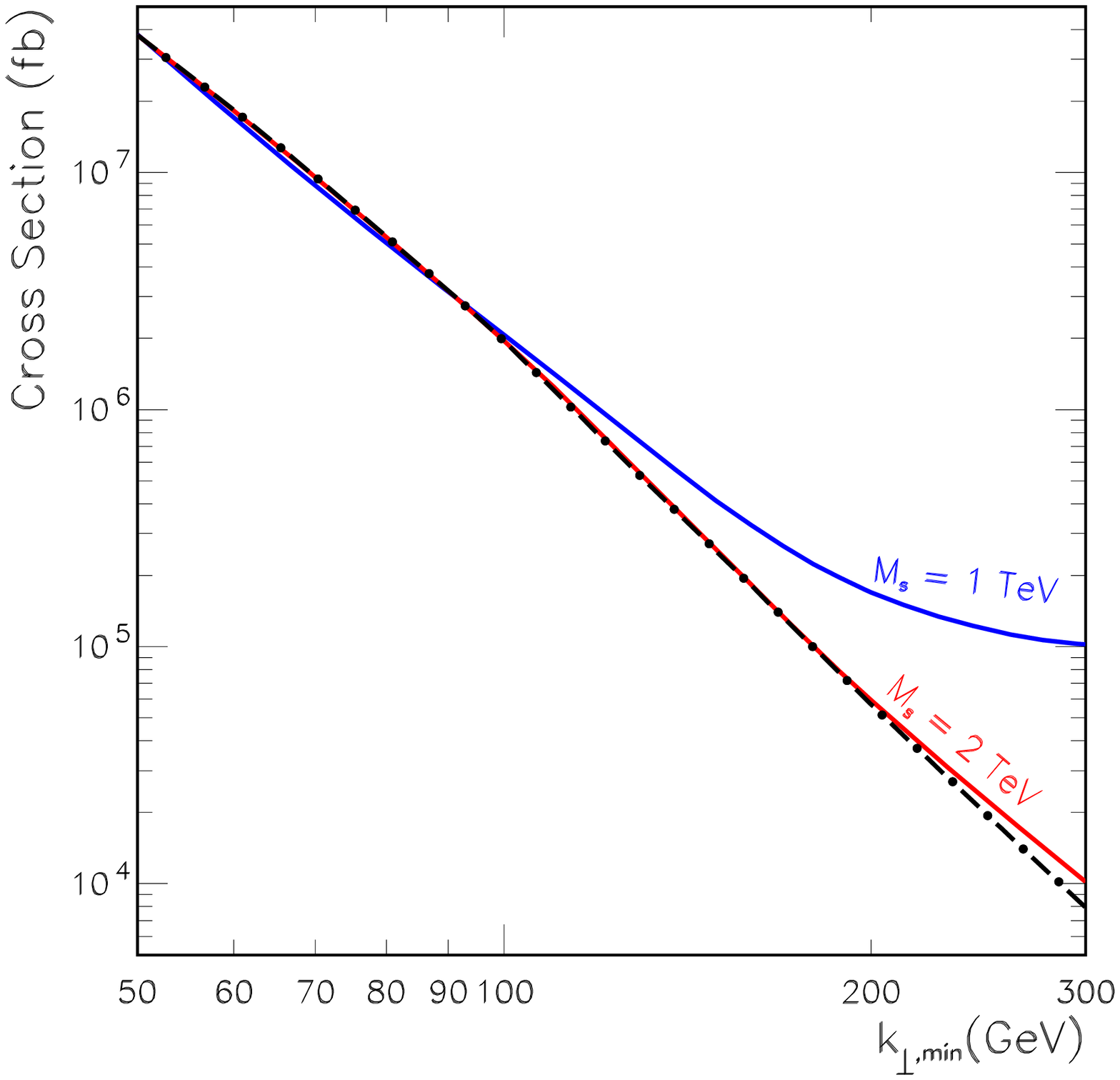}{0.99}
\end{minipage}
\hfill
\begin{minipage}[t]{0.49\textwidth}
\postscript{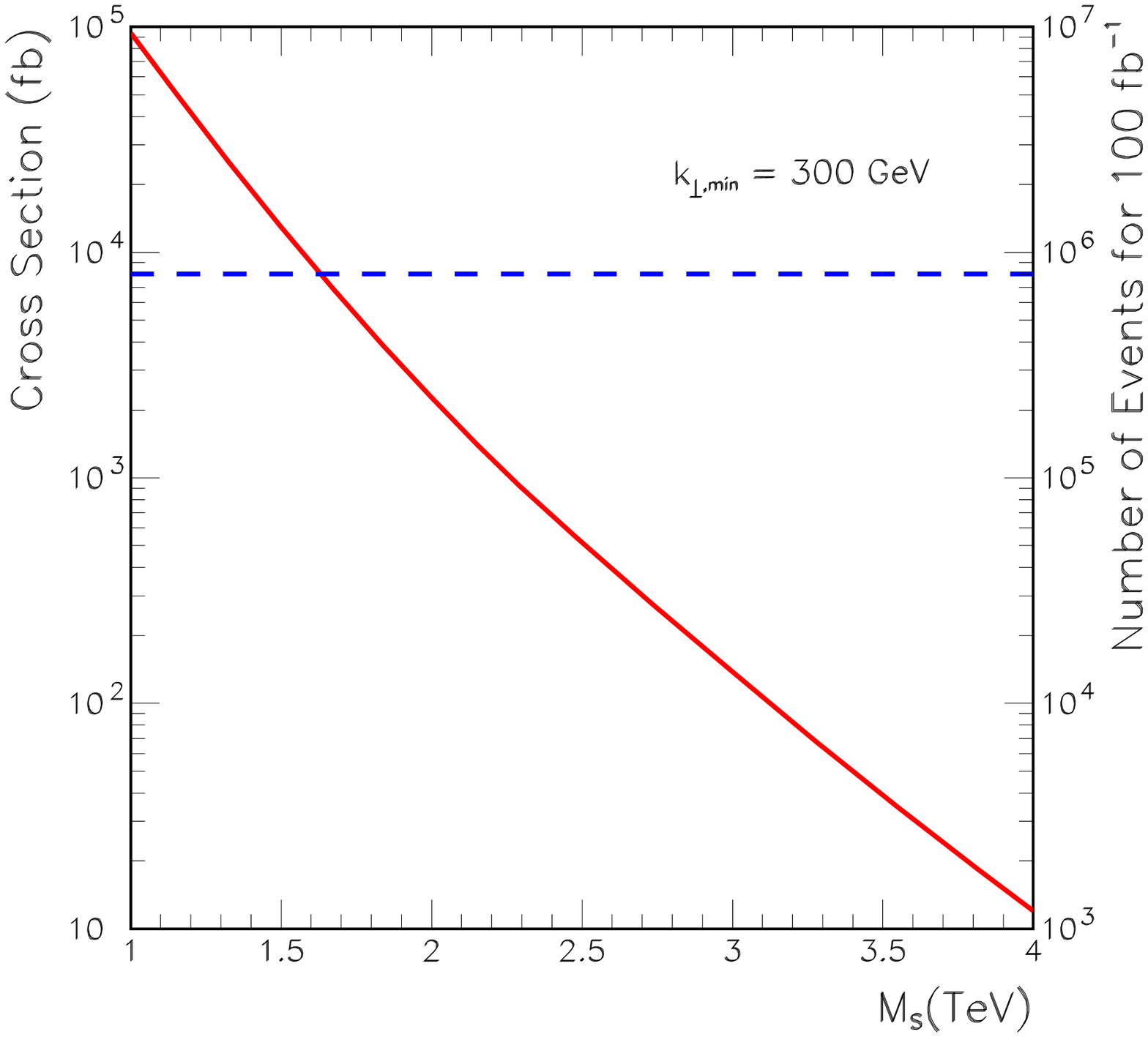}{0.99}
\end{minipage}
\caption{In the left panel we display the behavior of the QCD cross section
  (dot-dashed line) and string + QCD cross section (solid line) for $pp \to
  \gamma + {\rm jet}$ for two values of the string scale. In the right
  panel we show the cross section and the number of events for fixed 
  $k_{\perp{\rm min}} = 300$~GeV and varying string scale. The horizontal
  dashed line represents the SM background.}
\label{fig:sigma}
\end{figure}

In order to assess the possibility of discovery of new physics above
background at LHC, we adopt the kind of signal introduced
in~\cite{Dimopoulos:2001hw} to study detection of TeV-scale black
holes at the LHC, namely a high-$k_{\perp}$ isolated $\gamma$ or $Z.$
Thus, armed with parton distribution functions
(CTEQ6D)~\cite{Pumplin:2002vw} we have calculated integrated cross
sections $\sigma(pp\rightarrow \gamma + {\rm jet})|_{k_\perp(\gamma)>
  k_{\perp, {\rm min}}}$ for both the background QCD processes and for
$gg\rightarrow \gamma g$, for an array of values for the string
scale $M_s.$ Our results are shown in Fig.~\ref{fig:sigma}. As can
be seen in the left panel, the background is significantly reduced
for large $k_{\perp,{\rm min}}$. At very large values of $k_{\perp, {\rm min}},$
however, event rates become problematic. In the right panel we show
the cross section and number of events (before cuts) in a 100~fb$^{-1}$ 
run at LHC for both SM processes (dashed line) and for the
string amplitude (solid line), for $k_{\perp,{\rm min}}=300$~GeV, as a
function of the string scale $M_s$.

\begin{figure}
 \postscript{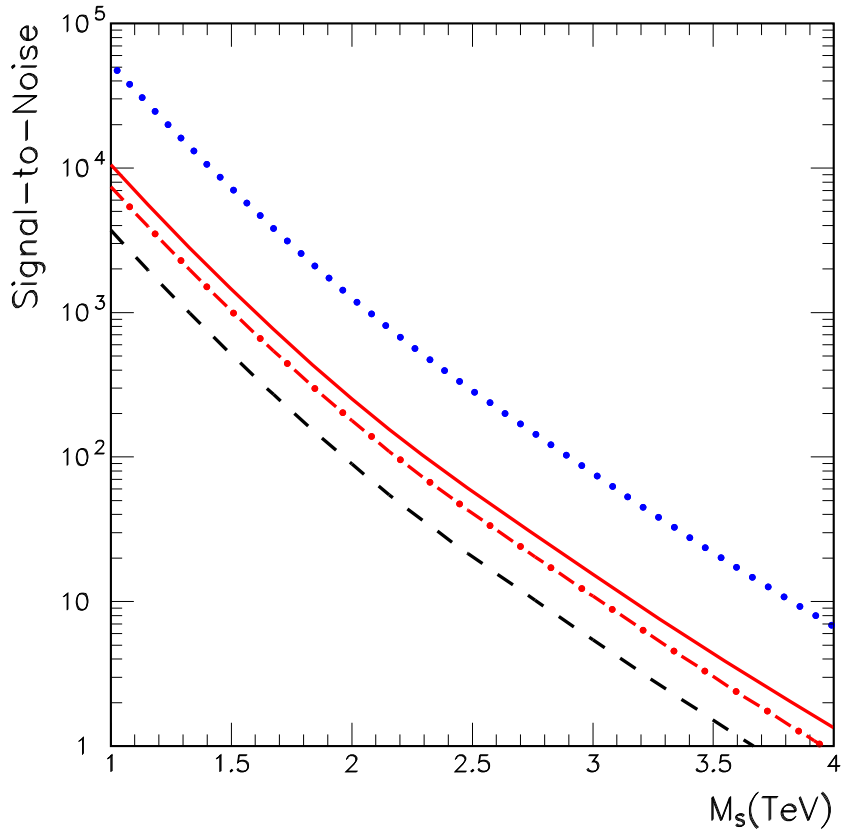}{0.7}
 \caption{Signal-to-noise ratio for an integrated luminosity of
   100~fb$^{-1}.$ The solid line indicates the optimistic case with
   100\% detector efficiency and $\kappa^2 = 0.02$.  The dashed
   ($\kappa^2 = 0.01$) and dot-dashed ($\kappa^2 = 0.02$) lines
   indicate more conservative scenarios in which considerations of
   detector efficiency and selection cuts reduce the total number of
   events by a factor of two. The dotted line corresponds to a very
   optimistic non-minimal case with $\kappa^2 = 0.1$ and 100\%
   detector efficiency.}
\label{fig:SN}
\end{figure}

Our significant results are encapsuled in Fig.~\ref{fig:SN}, where we
show the signal-to-noise ratio (${\rm signal}/\sqrt{\rm SM\
  background}$) as a function of $M_s$ for an integrated luminosity of
100~fb$^{-1}.$ The solid line indicates an optimistic case with
$\kappa^2 = 0.02$, and 100\% detector efficiency with no additional
cuts beyond $k_{\perp} (\gamma)>300$~GeV. This allows $5\sigma$
discovery for $M_s$ as large as 3.5~TeV. The dashed ($\kappa^2 =
0.01$) and dot-dashed ($\kappa^2 = 0.02$) lines indicate more
conservative scenarios in which considerations of detector efficiency
and $\gamma$ isolation cuts reduce the total number of events by a factor of 
two. In this case, for $\kappa^2=0.02$, discovery is
now possible for $M_s$ as large as 3.3~TeV. Even in the pessimistic
case, for $\kappa^2=0.01$ and 50\% detector efficiency, a string scale
as large as 3.1~TeV can be discovered. The dotted curve allows an         
illustrative view of the LHC reach in a hypothetical non-minimal (and           
optimistic) scenario in which the baryon $U(1)$ ($C_\mu$) has a 
sizeable hypercharge component. 

In summary, we have shown that cross section measurements of the
process $pp\rightarrow {\rm high-}k_{\perp}\ \gamma + {\rm jet}$  at the
LHC will attain 5$\sigma$ discovery reach on low scale  string
models for $M_{\rm string}$ as large as 3.3~TeV, even with detector
efficiency of 50\%~\cite{note}. In order to minimize misidentification with a
high-$k_{\perp}\ \pi^0$, isolation cuts must be imposed on the
photon, and to trigger on the desired channel, the hadronic jet must
be identified~\cite{Bandurin:2003kb}. We will leave the exact nature of
these cuts for the experimental groups.

In closing, we would like to note that the results presented here
are conservative, in the sense that we have not included in the
signal the stringy contributions to the SM processes. These will be
somewhat more model dependent since they require details of the
fermion quiver assignments, but we expect that these
contributions can potentially double the signal, significantly
increasing the reach of LHC for low-scale string discovery. In
addition, a similar treatment of $pp\rightarrow Z+ {\rm jet},\
Z\rightarrow \ell^+\ell^-$ could provide a potentially cleaner
signal. The stringy calculation to include transverse $Z$'s will
be presented in a future work.\\

L.A.A.\ is supported by the UWM Research Growth
Initiative.  H.G.\ is supported by the U.S. National Science
Foundation Grant No PHY-0244507.  The research of T.R.T.\ is supported
by the U.S. National Science Foundation Grant PHY-0600304 and by the
Cluster of Excellence "Origin and Structure of the Universe" in
Munich, Germany.  He is grateful to the participants of informal
string phenomenology seminars at Harvard University for illuminating
comments, and to Ignatios Antoniadis for useful remarks.  He is also
deeply indebted to Dieter L\"ust for a timely invitation to Munich.
Any opinions, findings, and conclusions or recommendations expressed
in this material are those of the authors and do not necessarily
reflect the views of the National Science Foundation.


\begin{thebibliography}{99}
\bibitem{nota} The width is adjusted to the higher spin of string
  excitations propagating in the $s$-channel.
\bibitem{note} Establishing the stringy origin of a resonance
  discovered at LHC is facilitated in the minimal model because of the
  simultaneous prencence of an intermediate mass $Z'$ with leptophobic 
  decay modes.
\end{thebibliography}
\end{document}